\documentstyle[aps,axodraw]{revtex}

\def\be{\begin{equation}}
\def\ee{\end{equation}}
\def\M{{\cal M}}
\def\bea{\begin{eqnarray}}
\def\eea{\end{eqnarray}}

\newcommand{\Section}[1]{\section{#1}\setcounter{equation}{0}}

\def\nn{\nonumber}

\def\o{\over}
\def\p{\partial}

\begin{document}
\title
{ Scalar Electrodynamics in Framework of Randall-Sundrum Model }
\author
{Amir Masoud Ghezelbash$^{1,2}$,
Kamran Kaviani $^{1,2}$,
and Shahrokh Parvizi $^{2}$}
\address
{\it $^1$ Department of Physics, Az-zahra University,
P.O.Box 19834, Tehran, Iran
\\
$^2$ Institute for Studies in Theoretical Physics and Mathematics (IPM),\\
P.O.Box 19395-5531, Tehran, Iran\\
E-mails:  amasoud,kaviani,parvizi@theory.ipm.ac.ir}
\maketitle
\vspace {-7cm}
\begin{flushright}
IPM/P2000/021  \\
hep-ph/0005166
\end{flushright}
\vspace*{6cm}
\begin{abstract}
Considering the Randall-Sundrum background,
we calculate the total cross-section for
$\phi\phi ^\star \rightarrow G \gamma$ in the framework of
the scalar electrodynamics.
\\
\end{abstract}
\Section{Introduction}
In \cite{rs1,rs2}, Randall and Sundrum (RS) proposed a new approach
to extra dimensions for space-time to solve the hierarchy
problem.
In their model, it is assumed that there exists only one extra
space-like dimension which is taken homeomorphic with an
orbifold $S^1 / Z_2$. This orbifold has
two fixed points at $\varphi=0$ and $\varphi=\pi$.
At each fixed point, they put a 4-dimensional brane world.
One of them which is located at $\varphi=\pi$
is called the visible world and is assumed we are living on it, and the
other one is called hidden world.

In RS method same as the Arkani-Hamed, Dimopoulos and Dvali (ADD) approach
\cite{anto,ark}, it is assumed that except the graviton (and also axions) all
the Standard Model (SM) fields are confined in these two distinct worlds. The
physical laws are the same on these two worlds but the masses and the
coupling constants may differ.

In this model, the classical action is assumed to be:
\bea
S&=& S_{gravity} + S_{vis} +S_{hid}, \nonumber \\
S_{gravity}&=& \int d^4x \int d \varphi \sqrt{-G} \{\Lambda + 2 M
R\}, \nonumber \\
S_{vis}&=& \int d^4x \sqrt{-g_{vis}} \{{\cal L}_{vis} -V_{vis}\}, \nonumber \\
S_{hid}&=& \int d^4x \sqrt{-g_{hid}} \{{\cal L}_{hid} -V_{hid}\},
\eea
where $M$ and $\Lambda$ are the 5-dimensional Planck mass and the
cosmological constant respectively, and ${\cal L}_{vis}$ (${\cal L}_{hid}$)
is the SM  or any effective Lagrangian corresponding to matter
and force fields except the gravity. The $V_{vis}$ and $V_{hid}$ are vacuum
expectations on the branes.

The classical solution of Einstein equation for the mentioned
action is the following metric:
\be
ds^2 = e^{-2 \sigma(\varphi)} \eta_{\mu\nu} dx^\mu dx^\nu + r_c^2 d\varphi^2,
\ee
with
\be
\sigma(\varphi) =\kappa r_c |\varphi|,
\ee
where $\varphi$ and $r_c$ are the coordinate and radius of $S^1$ in the orbifold
$S^1/Z_2$ and $\kappa =\sqrt{{-\Lambda\over {24 M^3}}}$.
By integrating out the fifth dimension, the coupling constant of the effective
4-dimensional action yields the 4-dimensional Planck scale \cite{rs1}:
\be
M_{P}={M^3 \o \kappa}(1-e^{-2\kappa r_c \pi}).
\ee
It is found that a field on visible brane with the fundamental mass parameter
$m_0$ will appear to have a physical mass $m=m_0e^{-\kappa r_c \pi}$.
By taking $\kappa r_c \simeq 12$, the observed scale hierarchy reproduces
naturally by exponential factor and no additional large hierarchies 
arise \cite{rs1,rs2}.

At this stage, it is natural to search for any observable effects of this
extra fifth dimension in RS model. Many efforts have been done to
probe the effects of this extra dimension in ordinary particle
interactions \cite{PES,dav1,dav2}.

There would be two kinds of gravitons in this formalism; the first type is
massless ordinary graviton, which is also confined to the 4-dimensional
physical space-time, and the others are massive gravitons. In \cite{dav1} it
is shown that the effects of the massless gravitons in particles
interactions are in order of $1\over {M_P^2}$ where $M_P$ is the 4-dim
Planck mass.
However, the contributions of the massive ones are considerable and
comparable with the weak scale of the Standard Model.
The masses of gravitons come from Kaluza-Klein compactification of the
fifth dimension. Due to non-factorization of the geometry, the masses
of gravitons are $m_n=\kappa x_n e^{-\sigma (\pi)}$, where $x_n$'s are
the roots of $J_1(x)$, the Bessel function of order one.

In this paper, we calculate the total cross-sections of scalar-scalar to
photon and graviton fields $(\phi\phi ^\star \rightarrow G \gamma)$.
This process in Standard Model, without producing gravitons is forbidden
by energy-momentum conservation.

\Section{Differential cross-section for $\phi \phi^\star \rightarrow G \gamma$}
At the beginning, we consider the scalar electrodynamics as the effective
theory of matter and forces in visible and hidden spaces. The action
of this theory is:

\bea
S_{vis}=S_{SED} &=& \int \{ g^{\mu\nu}(\p_\mu -i e A_\mu)\phi ^\star
(\p_\nu +i e A_\nu )\phi   \nn \\
&-&m_{\phi}^2 \phi \phi ^\star-{1 \over
4}g^{\alpha\mu}g^{\beta\nu}F_{\alpha\beta}F_{\mu\nu} \}\sqrt{-g} d^4x,
\eea
where $g^{\mu\nu}=g^{(0)\mu\nu} + \frac{1}{\Lambda}h^{\mu\nu} $ and
$g^{(0)\mu\nu}$ is the inverse of the classical metric (1.2).
The factor $\Lambda$ for massless graviton is equal to $M_P$ and for the
massive gravitons is $e^{-\sigma (\pi)} M_P$.
Inserting $g^{\mu\nu}$ in eq. (2.1) and absorbing the conformal factor
$e^{-2\sigma (\pi)}$ in scalar fields and their mass, one can reduce the
interaction part of Lagrangian up to the first order in $h^{\mu\nu}$ to the
following terms,
\bea \label{LAGRANGIAN}
{\cal L}_I&=& i e A^\mu (\p _\mu \phi ^\star \phi -\phi ^\star \p _\mu \phi )+
e^2 A^\mu A_\mu \phi ^\star \phi \nn\\
&+&\frac{h^{\mu\nu}}{\Lambda}\p _\mu \phi ^\star \p _\nu \phi +
i e \frac{h^{\mu\nu}}{\Lambda} A_\nu (\p _\mu \phi ^\star \phi -\phi ^\star
\p _\nu \phi )\nn\\
&-&
\frac{h^{\mu\nu}}{2\Lambda}(\p ^\lambda
A_\nu \p _\lambda A_\mu -2 \p ^\lambda A_\nu \p _\mu A_\lambda + \p _\nu
A^\lambda \p _\nu A_\lambda)\nn\\
&+&e^2 \frac{h^{\mu\nu}}{\Lambda} A_\mu A_\nu \phi ^\star \phi.
\eea
As it is pointed out in ref. \cite{ITZY}, we can use  $-{\cal L}_I$ as ${\cal
H}_I$, for this theory. Since we are searching for amplitude of $\phi \phi
^\star \rightarrow \gamma G$, the relevant terms of Hamiltonian to this
interaction are,
\bea \label{HAMILTON}
{\cal H}_1&=&-i e A^\mu (\p _\mu \phi ^\star \phi - \phi ^\star \p _\mu \phi
),\nn\\
{\cal H}_2&=&- \frac{h^{\mu\nu}}{\Lambda}\p _\mu \phi ^\star \p _\nu \phi, \nn\\
{\cal H}_3&=& -i e \frac{h^{\mu\nu}}{\Lambda} A_\nu (\p _\mu \phi ^\star \phi -
\phi ^\star \p
_\nu \phi ),\nn\\
{\cal H}_4&=&
\frac{h^{\mu\nu}}{2\Lambda}(\p ^\lambda
A_\nu \p _\lambda A_\mu -2 \p ^\lambda A_\nu \p _\mu A_\lambda + \p _\nu
A^\lambda \p _\nu A_\lambda),
\eea
which contribute to the following diagrams,
\vskip 1cm
\SetScale{0.5}
\begin{center}
\begin{picture}(0,0)(0,0)
\SetOffset(-150,-35)
\ArrowLine(-100,50)(0,0)
\ArrowLine(0,0)(-100,-50)
\Photon(0,0)(100,-50){5}{10}
\DashLine(0,0)(100,50){5}
\Vertex(0,0){3}
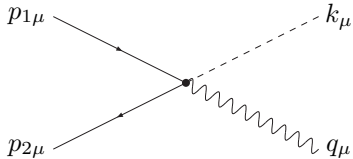
\Text(0,-67)[]{The Feynman Diagram of ${\cal M}_1$}
\Text(-60,25)[]{$p_{1\mu}$}
\Text(-60,-25)[]{$p_{2\mu}$}
\Text(58,25)[]{$k_\mu$}
\Text(58,-25)[]{$q_\mu$}
\SetOffset(150,-20)
\ArrowLine(-100,50)(0,0)
\DashLine(0,0)(100,50){5}
\Vertex(0,0){3}
\ArrowLine(0,0)(0,-65)
\ArrowLine(0,-65)(-100,-115)
\Photon(0,-65)(100,-115){5}{10}
\Vertex(0,-65){3}
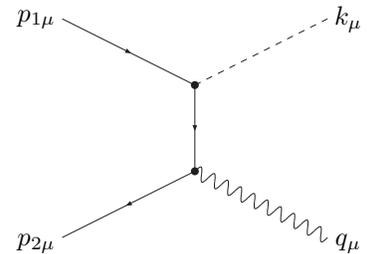
\Text(0,-83)[]{The Feynman Diagram of ${\cal M}_2$}
\Text(-60,25)[]{$p_{1\mu}$}
\Text(-60,-60)[]{$p_{2\mu}$}
\Text(58,25)[]{$k_\mu$}
\Text(58,-60)[]{$q_\mu$}
\end{picture}
\end{center}
\vskip 3.5cm
\newpage
\vspace*{.5cm}
\SetScale{0.5}
\begin{center}
\begin{picture}(0,0)(0,0)
\SetOffset(-150,0)
\ArrowLine(-100,50)(0,-65)
\ArrowLine(0,-65)(0,0)
\ArrowLine(0,0)(-100,-115)
\Vertex(0,0){3}
\Vertex(0,-65){3}
\DashLine(0,0)(100,50){5}
\Photon(0,-65)(100,-115){5}{10}
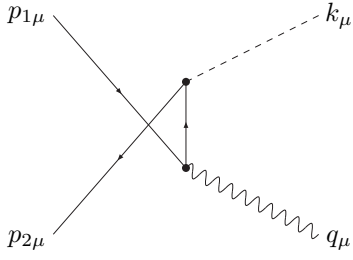
\Text(0,-83)[]{The Feynman Diagram of ${\cal M}_3$}
\Text(-60,25)[]{$p_{1\mu}$}
\Text(-60,-60)[]{$p_{2\mu}$}
\Text(58,25)[]{$k_\mu$}
\Text(58,-60)[]{$q_\mu$}
\SetOffset(115,-20)
\ArrowLine(-100,50)(0,0)
\ArrowLine(0,0)(-110,-50)
\Vertex(0,0){3}
\Vertex(105,0){3}
\DashLine(105,0)(205,50){5}
\Photon(0,0)(100,0){5}{10}
\Photon(105,0)(205,-50){5}{10}
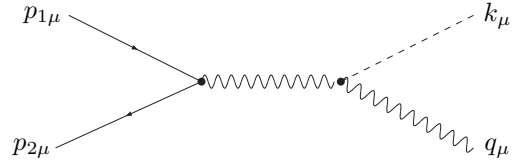
\Text(25,-60)[]{The Feynman Diagram of ${\cal M}_4$}
\Text(-60,25)[]{$p_{1\mu}$}
\Text(-64,-25)[]{$p_{2\mu}$}
\Text(112,25)[]{$k_\mu$}
\Text(112,-25)[]{$q_\mu$}
\end{picture}
\end{center}
\vspace{100pt}\hfill \\
Denoting in-state by $\mid p_1,p_2>$ where $p_1$ and $p_2$ are the momenta of
scalar and anti-scalar particles respectively and out-state by $\mid k,q>$,
where $k$ and $q$ are the momenta due to the massive graviton and photon.
Now, we are going to derive the $S$-matrix elements for the above
diagrams in the tree level,
\bea
<k,q \mid S \mid p_1 ,p_2>&=&<k,q \mid T e^{-i\int {\cal H}_I dt} \mid p_1
,p_2>\nn\\
&=&i (2\pi)^4 \delta ^{(4)} (q+k-p_1-p_2){\cal M}_{tot},
\eea
where ${\cal M}_{tot}$ is invariant amplitude which is the sum of the following
amplitudes (see figs.),
\bea \label{AMPL}
{\cal M}_1 &=& -\frac{e}{\Lambda} \epsilon _\mu (q) e^{\mu\nu} (k)
(p_1-p_2)_\mu,\nn\\
{\cal M}_2 &=& -\frac{e}{\Lambda}   \epsilon ^\lambda (q) e^{\mu\nu} (k)
{{(p_1-k)_\mu p_{1\nu} (p_2-p_1+k)_\lambda }\over {(p_1-k )^2+m_\phi ^2}},\nn\\
{\cal M}_3 &=& -\frac{e}{\Lambda}   \epsilon ^\lambda (q) e^{\mu\nu} (k)
{{(p_2-k)_\mu p_{2\nu} (p_2-p_1-k)_\lambda }\over {(p_2-k )^2+m_\phi ^2}},\nn\\
{\cal M}_4&=&\frac{e}{\Lambda}  {{e^{\mu\nu} (k) }\over {(p_1+p_2)^2+i\epsilon}}
\left(
(p_1-p_2)_\nu \epsilon _\mu (q\cdot (p_1+p_2))\right.\nn\\
&-&\left.(p_1-p_2)_\nu q_\mu (\epsilon
\cdot (p_1+p_2))-\epsilon _\nu (p_1+p_2)_\mu (p_1-p_2)\cdot q\right.\nn\\
&+&\left.q_\nu (p_1+p_2)_\mu (p_1-p_2)\cdot \epsilon \right).
\eea
In the above equation $\epsilon^\mu$ and $e^{\mu\nu}$ are the polarization
of photon and graviton respectively.
To calculate the unpolarized cross-section, we should make summation over these
polarizations. We have,
\bea
\sum
_{pol.}e_{\mu\nu}(k)e_{\alpha\beta}(k)&=&f_{\mu\nu\alpha\beta}(k),\nonumber\\
\sum
_{pol.}\epsilon_{\mu}(q)\epsilon_{\nu}(q)&=&-g_{\mu\nu},
\eea
in which $f_{\mu\nu\alpha\beta}$ for a massive graviton is \cite{VELT},
\bea
& &f_{\mu\nu\alpha\beta}(k)={1\over 2}\left\{
g_{\mu\alpha}g_{\nu\beta}+g_{\mu\beta}g_{\nu\alpha}
-g_{\mu\nu}g_{\alpha\beta}\right\}        \nonumber\\
&+&{1\over 2}\left\{
g_{\mu\alpha}{{k_\nu k_\beta}\over {m^2}}+
g_{\nu\beta}{{k_\mu k_\alpha}\over {m^2}}+
g_{\mu\beta}{{k_\nu k_\alpha}\over {m^2}}+
g_{\nu\alpha}{{k_\mu k_\beta}\over {m^2}}
\right\}                                   \nonumber\\
&+&{2 \over 3}({1 \over 2}g_{\mu\nu}-{{k_\mu k_\nu}\over {m^2}})
({1 \over 2}g_{\alpha\beta}-{{k_\alpha k_\beta}\over {m^2}}),
\eea
and for a massless graviton,
\be
f_{\mu\nu\alpha\beta}(k)={1\over 2}\left\{
g_{\mu\alpha}g_{\nu\beta}+g_{\mu\beta}g_{\nu\alpha}
-{2\over 3}g_{\mu\nu}g_{\alpha\beta}\right\}.
\ee
Using the above relations,  it is straightforward to calculate the
cross-section in the center of mass frame of the incident
particles for the massive  gravitons.
The differential cross-section of the mentioned process in the
center of mass frame is \cite{CROS},
\be \label{CS}
({{d\Sigma}\over{d\Omega}})_{CM}=
{1 \over {256 \pi ^2 E^2}} {{|\vec k|}\over {|\vec p_1|}} |{\cal M}_{tot}|^2,
\ee
where ${\cal M}_{tot}={\cal M}_{1}+{\cal M}_{2}+{\cal M}_{3}+{\cal M}_{4}$,
and $2E$ is the center of mass energy of incident particles.
The contribution of the
massless graviton can be neglected, as it is pointed out in \cite{dav1}.
For a massive graviton of mass $m$, we have calculated in the Appendix
the terms in
$\sum _{pol.} |{\cal M}_{tot}|^2$.
To obtain the total cross-section, as a function of $E$,
one should integrate the
(\ref{CS}) over the scattering angles and sum over all massive gravitons
which their masses are less than $2E$.
For the calculations the computer program Mathematica version 3.0 was used.

The final result is the following graph, where
the solid curve shows the behavior of the total cross-section versus the
energy, $E$. Here, we take $m_{\phi}=0.2\,TeV$, $\sigma(\pi)=12\pi$ and
$\kappa=10^{16}\,TeV$.
According to the mass formula, $m_n=\kappa x_n \exp {(-12\pi)}$, we obtain
the first four gravitons' masses, $m_1=1.6\,TeV$, $m_2=2.9\,TeV$,
$m_3=4.2\,TeV$ and $m_4=5.6\,TeV$.

For $E < 1.45\,TeV$, only the first graviton mode contributes
to the total cross-section. For $E \geq 1.45\, TeV$, the dashed curve shows
the contribution of this first mode to the $\Sigma _{tot}$.
The individual behavior of the other graviton modes are similar to this
dashed curve which shows a monotonic increasing behavior.
This increasing behavior is expected due to the non-renormalizability of the
quantum gravity. The peaks on the solid curve show the resonance behavior
according to creation of the graviton modes.

\newpage
\begin{center}
\begin{picture}(0,0)(0,0)
\SetOffset(0,0)
\SetScale{0.35}
\LinAxis(-300,0)(300,0)(10,5,-4,0,1.5)
\LinAxis(-300,-700)(300,-700)(10,5,4,0,1.5)
\LinAxis(-300,0)(-300,-700)(12,5,4,0,1.5)
\LinAxis(300,0)(300,-700)(12,5,-4,0,1.5)
\Text(-105,-252)[]{\tiny .8}
\Text(-84,-252)[]{\tiny 1}
\Text(-63,-252)[]{\tiny 1.2}
\Text(-42,-252)[]{\tiny 1.4}
\Text(-21,-252)[]{\tiny 1.6}
\Text(0,-252)[]{\tiny 1.8}
\Text(21,-252)[]{\tiny 2}
\Text(42,-252)[]{\tiny 2.2}
\Text(63,-252)[]{\tiny 2.4}
\Text(84,-252)[]{\tiny 2.6}
\Text(105,-252)[]{\tiny 2.8}
\Text(0,-270)[]{$E$ in TeV}
\rText(-130,-126)[][l]{\Large $\Sigma$ in pb}
\SetWidth{3}
\Curve{(-296.25,-17.2759)(-270,-607.7608)(-255,-647.5)
(-247.5,-653.4628)(-243.75,-655.7726)
(-240,-657.6917)(-210,-665.0621)
(-180,-666.1088)
(-150,-664.7961)(-135,-663.0000)(-130,-662.0000)(-125,-662.1000)
(-120,-662.3027)
(-90,-17.2842)(-60,-502.8217)(-45,-550.0000)(-30,-567.1108)(0,-587.1512)
(30,-593.4198)
(60,-593.5084)(75,-590.0000)(82,-590.1200)(86,-590.1500)
(90,-590.1525)(119,-17.2792)(150,-374.4551)(180,-445.7874)
(210,-471.2680)(240,-479.3320)(270,-478.6198)(300,-472.6167)}
\DashCurve{(-120,-662.3027)(-90,-659.0552)(-60,-655.2215)(-30,-650.8643)
(0,-645.9984)(30,-640.6155)(60,-634.6940)(90,-628.2045)
(120,-621.1135)(150,-613.3826)(180,-604.9702)(210,-595.8359)
(240,-585.9309)(270,-575.2098)(300,-563.6319)}{10}
\Text(-112,-245)[r]{\tiny 0} \Text(-112,-225)[r]{\tiny 10}
\Text(-112,-205)[r]{\tiny 20}
\Text(-112,-185)[r]{\tiny 30}
\Text(-112,-164.4)[r]{\tiny 40}
\Text(-112,-143.8)[r]{\tiny 50}
\Text(-112,-123.2)[r]{\tiny 60}
\Text(-112,-102.6)[r]{\tiny 70}
\Text(-112,-82)[r]{\tiny 80}
\Text(-112,-61.4)[r]{\tiny 90}
\Text(-112,-40.8)[r]{\tiny 100}
\Text(-112,-20.2)[r]{\tiny 110}
\Text(-112,0)[r]{\tiny 120}
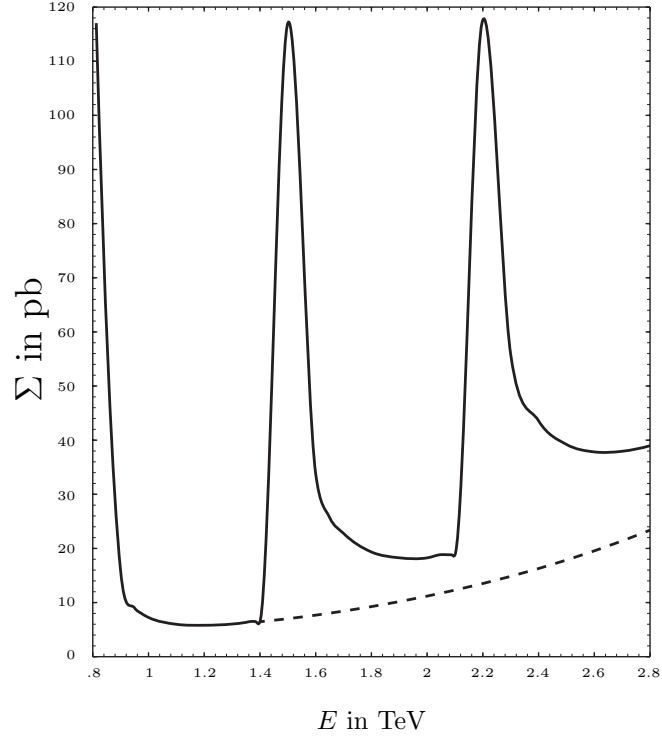
\Text(0,-295)[]{Total cross section of $\phi \phi ^\star \rightarrow G \gamma$
versus energy of one incident particles, with }
\Text(0,-310)[]{$m_{\phi}=0.2 TeV$. The peaks correspond to the
productions of massive gravitons.}
\end{picture} \end{center}
\vspace{10cm}
\vspace*{1cm}
\section*{Acknowledgments}
We would like to thank A. Shafiekhani for his contribution to the primary
version of this article.
\vspace{1cm}
\section*{Appendix}
In this appendix, all the ten terms of the scattering diagrams \hspace {7mm}
$
\SetScale{0.2}
\begin{picture}(0,0)(0,0)
\ArrowLine(-100,50)(0,0)
\ArrowLine(0,0)(-100,-50)
\Photon(0,0)(100,-50){5}{10}
\DashLine(0,0)(100,50){5}
\CCirc(0,0){50}{4}{0}
\end{picture}
$ \hspace {7mm}
up to the order of ${e \over {M_p}}$  for $m_{\phi}=0.2\, TeV$ in terms of
the mass of graviton, $m$, and enrgy, $E$, have been calculated.
\bea
\frac{\Lambda ^2}{e^2}\sum_{pol.}|{\cal M}_1|^2&=&
{\frac{5\,\left( 4\,{E^2} - m^2 \right) \,
     \left( 4\,{c^2} + \left( -0.16 + 4\,{E^2} \right) \,m^2 \right) }{3072
     \,{E^3}\,
     {\sqrt{-0.04 + {E^2}}}\,m^2\,{{\pi }^2}}},
\nn\\
\frac{\Lambda ^2}{e^2}\sum_{pol.}(\M_1 \M _2^\star  &+&\M _1 ^\star \M _2)=
\frac{-1}
{3072\,{E^3}\,{\sqrt{-0.04 + {E^2}}}\,\left( 4\,{E^2} - 4\,c - m^2 \right) \,
     {m^4}\,{{\pi }^2}}
   \left( \left( 4\,{E^2} - m^2 \right) \,
\left( 32.\,{c^4} - 16.\,{E^6}\,m^2
\right.\right.
\nn\\
&-&\left.\left.   16.\,{c^3}\,m^2 + 0.64\,c\,{m^4} +
\left( 0.0512 + 0.04\,m^2 \right) \,{m^4}
+ {c^2}\,m^2\,\left( -2.56 + 2.\,m^2 \right)  +
{E^2}\,\left( -64.\,{c^3} + 96.\,{c^2}\,m^2 +
\right.\right.\right.
\nn\\
&+& \left.\left.\left.
c\,\left( 2.56 - 16.\,m^2 \right) \,m^2 +
\left( -2.88 - 1.\,m^2 \right) \,{m^4} \right)  +
{E^4}\,\left( 32.\,{c^2} - 64.\,c\,m^2
+ m^2\,\left( 0.64 + 40.\,m^2 \right)  \right)  \right)
\right),
       \nn\\
\frac{\Lambda ^2}{e^2}\sum_{pol.}(\M_1 \M _3^\star  &+&\M _1 ^\star \M _3)=
\frac{1}
       {  {E^3}\,
     {\sqrt{-0.04 + {E^2}}}\,\left( 4\,{E^2} + 4\,c - m^2 \right) \,{m^4}
       }
\left( 0.000527714\,\left( 4\,{E^2} - m^2 \right) \,
           \left( -2.\,{c^4} + 1.\,{E^6}\,m^2
           \right.\right.
           \nn\\
           &-& \,{c^3}\,m^2
           +{c^2}\,\left( 0.16 - 0.125\,m^2 \right) \,m^2 + 0.04\,c\,{m^4}
      +
\left( -0.0032 - 0.0025\,m^2 \right) \,{m^4} +
       {E^4}\,\left( -2.\,{c^2} - 4.\,c\,m^2
       \right.
       \nn\\
       &+&\left.\left.\left. \left( -0.04 - 2.5\,m^2 \right) \,m^2 \right)  +
       {E^2}\,\left( -4.\,{c^3} - 6.\,{c^2}\,m^2
       + c\,\left( 0.16 - 1.\,m^2 \right) \,m^2 +
          \left( 0.18 + 0.0625\,m^2 \right) \,{m^4} \right)  \right)
       \right),
 \nn\\
\frac{\Lambda ^2}{e^2}\sum_{pol.}(\M_1 \M _4^\star &+&\M _1 ^\star \M _4)=
{\frac{E\,\left( 0.0000527714 + 0.000989465\,m^2 \right) }{{\sqrt{-0.04
+ {E^2}}}\,m^2}} -
  {\frac{8.24554\,{{10}^{-7}}\,{m^4}}{{E^7}\,{\sqrt{-0.04 + {E^2}}}}}
   - {\frac{5\,{E^3}}{384\,{\sqrt{-0.04 + {E^2}}}\,m^2\,{{\pi }^2}}}
  \nn\\
&+& {\frac{{c^2}\,\left( 0.000659643 - 0.000329822\,m^2 \right)  +
      0.0000206138\,\left( -7.60416 + m^2 \right) \,m^2\,\left( 0.0841644
      + m^2 \right) }{{E^3}\,
      {\sqrt{-0.04 + {E^2}}}\,m^2}}
  \nn\\
  &+&
  {\frac{0.00131929\,{c^2} + \left( 0.000290243 - 0.000247366\,m^2 \right)
  \,m^2}
    {E\,{\sqrt{-0.04 + {E^2}}}\,m^2}} \nn\\
    &+& {\frac{-0.000164911\,{c^2} +
      \left( 6.59643\,{{10}^{-6}} + 0.0000197893\,m^2 \right) \,m^2}{{E^5}\,
      {\sqrt{-0.04 + {E^2}}}}},
\nn\\
\frac{\Lambda ^2}{e^2}\sum_{pol.}|{\cal M}_2|^2&=&
\frac{1}
{98304\,{E^3}\,{\sqrt{-0.04 + {E^2}}}\,{m^4}\,{{\left( -4\,{E^2} + 4\,c
+ m^2 \right) }^2}\,
{{\pi }^2}
}
\left(4096.\,{E^{12}} + {E^{10}}\,\left( -163.84 - 20480.\,c
+ 2048.\,m^2 \right)\right.
\nn\\
&+&
     {E^8}\,\left( 40960.\,{c^2} + c\,\left( 655.36 - 7168.\,m^2 \right)  +
        \left( -450.56 - 256.\,m^2 \right) \,m^2 \right)
        \nn\\
       &+&
     {E^4}\,\left( 20480.\,{c^4} + {c^3}\,\left( 655.36 - 2048.\,m^2 \right)+
        {c^2}\,\left( -1228.8 - 3072.\,m^2 \right) \,m^2 \right.
        \nn\\
        &+&\left.
640.\,c\,\left( -0.276264 + m^2 \right) \,m^2\,
\left( 0.148264 + m^2 \right)  -
 16.\,\left( -3.07951 + m^2 \right) \,{m^4}\,\left( 0.199512 + m^2 \right)
 \right)  \nn\\
 &+&
 {E^2}\,\left( -4096.\,{c^5} + {c^4}\,\left( -163.84 - 2048.\,m^2 \right)  -
 512.\,{c^2}\,\left( -0.492029 + m^2 \right) \,m^2\,\left( 0.0520294
 + m^2 \right)\right. \nn\\
 &+&
 8.\,\left( -0.64 + m^2 \right) \,{m^4}\,\left( 0.0626408 + m^2 \right) \,
         \left( 0.817359 + m^2 \right)  -
 16.\,c\,{m^4}\,\left( 0.0812907 + m^2 \right) \,\nn\\
 &\times&\left.
 \left( 5.03871 + m^2 \right)
 +
 {c^3}\,m^2\,\left( 327.68 + 2560.\,m^2 \right)  \right)  +
 {E^6}\,\left( -40960.\,{c^3} - 256.\,\left( -0.189783 + m^2 \right) \,m^2\,
 \right. \nn\\
 &\times&\left.
 \left( 0.269783 + m^2 \right)  + c\,m^2\,\left( 1310.72 + 1536.\,m^2 \right)
 +
 {c^2}\,\left( -983.04 + 8192.\,m^2 \right)  \right)  +
  m^2\,\left( 1024.\,{c^5} \right.
  \nn\\
  &+& {c^4}\,\left( 40.96 - 768.\,m^2 \right)  -
 12.\,c\,\left( -0.64 + m^2 \right) \,{m^4}\,\left( 0.426667 + m^2 \right)  +
  32.\,{c^2}\,\left( -0.0849806 + m^2 \right) \,m^2\,\nn\\
  &\times&\left.\left.
  \left( 1.20498 + m^2 \right)  +
    {c^3}\,m^2\,\left( -122.88 + 128.\,m^2 \right)  +
 1.\,{m^4}\,\left( 0.16 + m^2 \right)
 \,\left( 0.4096 - 1.28\,m^2 + {m^4} \right)
 \right) \right),
 \nn\\
\frac{\Lambda ^2}{e^2}\sum_{pol.}(\M_2 \M _3^\star &+&\M _2 ^\star \M _3)=
\frac{\left( 4\,{E^2} - m^2 \right) \,\left( -0.16 + 4\,{E^2} + m^2 \right)}
{49152\,
   {E^3}\,{\sqrt{-0.04 + {E^2}}}\,{m^4}\,
\left( 16\,{E^4} - 16\,{c^2} - 8\,{E^2}\,m^2 + {m^4} \right) \,{{\pi }^2}}
\nn\\
&\times&
     \left( 256.\,{E^8} + 256.\,{c^4} - 1280.\,{E^6}\,m^2 +
       {c^2}\,\left( -20.48 - 32.\,m^2 \right) \,m^2 +
1.\,{m^4}\,\left( 0.171487 + m^2 \right) \, \right.   \nn\\
&\times&\left( 2.38851 + m^2 \right)  +
{E^2}\,m^2\,\left( 1280.\,{c^2} + \left( -40.96 - 80.\,m^2 \right)
\,m^2 \right)  \nn\\
        &+&\left.
{E^4}\,\left( -512.\,{c^2} + m^2\,\left( 40.96 + 864.\,m^2
\right)  \right)  \right),
       \nn\\
\frac{\Lambda ^2}{e^2}\sum_{pol.}(\M_2 \M _4^\star &+&\M _2 ^\star \M _4)=
\frac{1}
{24576\,{E^5}\,{\sqrt{-0.04 + {E^2}}}\,
\left( 4\,{E^2} - 4\,c - m^2 \right) \,{m^4}\,{{\pi }^2}}
\left(-1024.\,{E^{12}} - 7.10543\,{{10}^{-15}}\,{c^4}\,{m^2}\right.
\nn\\
&+&
   3.55271\,{{10}^{-15}}\,{c^3}\,{m^6} +
{c^2}\,\left( -1.92 - 4.44089\,{{10}^{-16}}\,m^2 \right) \,
{m^6}  +
     \left( 0.0384 + 0.12\,m^2 \right) \,{m^8} + \nn\\
&+&
    {E^{10}}\,\left( 40.96 + 2048.\,c + 2304.\,m^2 \right)  +
{E^8}\,\left( 1.13687\,{{10}^{-13}}\,{c^2}
+ c\,\left( -81.92 - 4096.\,m^2 \right) \right. \nn\\
&+&\left.
        \left( -112.64 - 640.\,m^2 \right) \,m^2 \right)  +
{E^4}\,\left( 1024.\,{c^4} + 1024.\,{c^3}\,m^2
+ {c^2}\,\left( -30.72 - 512.\,m^2 \right) \,m^2 \right. \nn\\   &+&
\left. c\,\left( -35.84 - 128.\,m^2 \right) \,{m^4} +
44.\,{m^4}\,\left( 0.0220983 + m^2 \right) \,
\left( 0.210629 + m^2 \right)  \right)  \nn\\    &+&
     {E^2}\,m^2\,\left( -256.\,{c^4} - 128.\,{c^3}\,m^2 -
   3.\,{m^4}\,\left( 0.106667 + m^2 \right) \,\left( 0.8 + m^2 \right)  +
c\,{m^4}\,\left( 1.28 + 8.\,m^2 \right) \right. \nn\\   &+& \left. {c^2}\,m^2
\,\left( 12.8 + 64.\,m^2 \right)  \right)
+ {E^6}\,\left( -2048.\,{c^3} - 96.\,\left( -0.247773 + m^2 \right) \,m^2\,
\left( 0.0344401 + m^2 \right)\right.  \nn\\
&+&\left.\left. {c^2}\,\left( 40.96 + 1024.\,m^2 \right)
+         c\,m^2\,\left( 143.36 + 1280.\,m^2 \right)  \right) \right),
     \nn\\
\frac{\Lambda ^2}{e^2}\sum_{pol.}|{\cal M}_3|^2&=&
\frac{1}
{98304\,{E^3}\,{\sqrt{-0.04 + {E^2}}}\,{m^4}\,
{{\left( -4\,{E^2} - 4\,c + m^2 \right) }^2}\,
{{\pi }^2}}
\left(4096.\,{E^{12}} + {E^{10}}\,\left( -163.84 + 20480.\,c
\right.\right.
\nn\\
&+&\left.2048.
\,m^2 \right)  +
     {E^2}\,\left( 4096.\,{c^5} + {c^4}\,\left( -163.84
 -2048.\,m^2 \right)
     +
        {c^3}\,\left( -327.68 - 2560.\,m^2 \right) \,m^2
\right.\nn\\
       &-&
        512.\,{c^2}\,\left( -0.492029 + m^2 \right) \,m^2\,
        \left( 0.0520294 + m^2 \right)  +
        8.\,\left( -0.64 + m^2 \right) \,{m^4}\,
        \left(  0.0626408 + m^2 \right) \,
\nn\\
&\times &\left.\left( 0.817359 + m^2 \right)  +
16.\,c\,{m^4}\,\left( 0.0812907 + m^2 \right) \,
\left( 5.03871 + m^2 \right)  \right)  +
 {E^4}\,\left( 20480.\,{c^4}\right.
\nn\\
&+&{c^2}\,
 \left( -1228.8 - 3072.\,m^2 \right) \,m^2 -
 640.\,c\,\left( -0.276264 + m^2 \right) \,m^2\,\left( 0.148264
 + m^2 \right)
 -
  16.\,
\nn\\
&\times&\left.  \left( -3.07951 + m^2 \right) \,{m^4}\,\left( 0.199512
  + m^2 \right)  +
  {c^3}\,\left( -655.36 + 2048.\,m^2 \right)  \right)  +
 {E^8}\,\left( 40960.\,{c^2} \right.
\nn\\
&-&\left.\left. \left(450.56+ 256.\,m^2 \right) \,m^2 +
  c\,\left( -655.36 + 7168.\,m^2 \right)  \right)  +
{E^6}\,\left( 40960.\,{c^3} - c\,\left(1310.72+1536.\,m^2 \right) \,m^2
\right.\right.
\nn\\
&-&\left.\left.
256.\,\left( -0.189783 + m^2 \right) \,m^2\,\left( 0.269783 + m^2 \right)  +
 {c^2}\,\left( -983.04 + 8192.\,m^2 \right)  \right)  +
  m^2\,\left( -1024.\,{c^5} \right.\right.
\nn\\
&+& {c^4}\,\left( 40.96 - 768.\,m^2 \right)  +
   {c^3}\,\left( 122.88 - 128.\,m^2 \right) \,m^2 +
   12.\,c\,\left( -0.64 + m^2 \right) \,{m^4}\,\left( 0.426667 + m^2
   \right)
   \nn\\
   &+&\left.\left.
  32.\,{c^2}\,\left( -0.0849806 + m^2 \right) \,m^2\,\left( 1.20498
  + m^2 \right)  +
   1.\,{m^4}\,\left( 0.16 + m^2 \right) \,\left( 0.4096 - 1.28\,m^2
  + {m^4} \right)  \right) \right),
\nn\\
\frac{\Lambda ^2}{e^2}\sum_{pol.}(\M_3 \M _4^\star &+&\M _3 ^\star \M _4)=
\frac{
\left( 0.16 - 4\,{E^2} - 2\,c \right) \,\left( E - {\frac{m^2}{4\,E}}
 \right)}
{512\,{E^4}\,
 {\sqrt{-0.04 + {E^2}}}\,\left( 4\,{E^2} + 4\,c - m^2 \right) \,
 {{\pi }^2}}
   \left( {\frac{{c^2}}{3}} + {E^4}\,
 \left( 6 + {\frac{16\,{c^2}}{3\,{m^4}}}- {\frac{8\,c}{3\,m^2}}
 \right) \right.
 \nn\\
 &+&\left.
 {E^2}\,\left( {\frac{-2\,c}{3}} + {\frac{16\,{c^2}}{3\,m^2}}
 - {\frac{11\,m^2}{6}} \right)  +
 {\frac{16\,{E^8}}{3\,{m^4}}}
 + {\frac{8\,{E^6}\,\left( 4\,c - 3\,m^2 \right) }{3\,{m^4}}} +
 {\frac{c\,m^2}{6}} + {\frac{3\,{m^4}}{16}} \right)
 \nn\\
&-&
  \frac{1}
      {3072\,{E^5}\,{\sqrt{-0.04 + {E^2}}}\,\left( 4\,{E^2} + 4\,c
      - m^2 \right) \,{m^4}\,
      {{\pi }^2}}
 \left(c\,\left( 4\,{E^2} - m^2 \right) \,
   \left( -8\,{E^6}\,\left( 2\,c + 3\,m^2 \right)  +
  m^2\,\left( -4.\,{c^3} \right.\right.\right.
\nn\\
&-& \left.
2.\,{c^2}\,m^2 + c\,
  \left( 0.24 - 2.25\,m^2 \right) \,m^2 +
      0.06\,{m^4} \right)  + {E^2}\,
   \left( -16.\,{c^3} + 8.\,{c^2}\,m^2 + c\,
   \left( 0.96 - 11.\,m^2 \right) \,m^2\right.
   \nn\\
   &+&\left.\left.\left.
        \left( -0.48 - 1.5\,m^2 \right) \,{m^4} \right)  +
 {E^4}\,\left( -32.\,{c^2}
 - 12.\,c\,m^2 + m^2\,\left( 0.96 + 12.\,m^2 \right)  \right)  \right)
\right)
\nn\\
&-& \frac{1}
{12288\,{E^5}\,{\sqrt{-0.04 + {E^2}}}\,\left( 4\,{E^2} + 4\,c
- m^2 \right) \,{m^4}\,
      {{\pi }^2}}
\left(\left( 4\,{E^2} - m^2 \right) \,\left( 4\,{E^2} + m^2 \right) \,
      \left( -8\,{E^6}\,\left( 2\,c + 3\,m^2 \right)  \right.\right.
 \nn\\
      &+&
        m^2\,\left( -4.\,{c^3} - 2.\,{c^2}\,m^2 + c\,\left( 0.24 -
        2.25\,m^2 \right) \,m^2 +
           0.06\,{m^4} \right)  + {E^2}\,
         \left( -16.\,{c^3} + 8.\,{c^2}\,m^2 \right.
  \nn\\
   &+& \left.\left.\left. c\,
         \left( 0.96 - 11.\,m^2 \right) \,m^2 +
           \left( -0.48 - 1.5\,m^2 \right) \,{m^4} \right)  +
        {E^4}\,\left( -32.\,{c^2} - 12.\,c\,m^2 + m^2\,
        \left( 0.96 + 12.\,m^2 \right)  \right)  \right) \right) +
\nn\\
&+& \frac{1}{24576\,{E^5}\,{\sqrt{-0.04 + {E^2}}}\,\left( 4\,{E^2} +
      4\,c - m^2 \right) \,{m^4}\,
      {{\pi }^2}}
\left({{\left( -4\,{E^2} + m^2 \right) }^2}\,
      \left( 32.\,{c^4} - 16.\,{E^6}\,m^2 + 16.\,{c^3}\,m^2 \right.\right.
      \nn\\
     &-&  0.64\,c\,{m^4} +
        \left( 0.0768 - 0.6\,m^2 \right) \,{m^4} + {c^2}\,m^2\,
        \left( -3.2 + 18.\,m^2 \right)  +
        {E^2}\,\left( 64.\,{c^3} + 96.\,{c^2}\,m^2 \right.
    \nn\\
        &+& \left.\left.\left.{m^4}\,
        \left( -3.52 + 15.\,m^2 \right)  +
           c\,m^2\,\left( -2.56 + 16.\,m^2 \right)  \right)  +
        {E^4}\,\left( 32.\,{c^2} + 64.\,c\,m^2 + m^2\,
        \left( 0.64 + 40.\,m^2 \right)  \right)  \right) \right),
\nn\\
\frac{\Lambda ^2}{e^2}\sum_{pol.}|{\cal M}_4|^2&=&
  {\frac{5\,{{\left( 4\,{E^2} - m^2 \right) }^3}\,
      \left( 4\,{c^2} + \left( -0.16 + 4\,{E^2} \right) \,m^2 \right)
      }{786432\,{E^7}\,
      {\sqrt{-0.04 + {E^2}}}\,m^2\,{{\pi }^2}}} +
  {\frac{5\,{c^2}\,{{\left( -4\,{E^2} + m^2 \right) }^2}\,
  \left( 4\,{E^2} + m^2 \right) }
    {98304\,{E^7}\,{\sqrt{-0.04 + {E^2}}}\,m^2\,{{\pi }^2}}}
  \nn\\
   &+&
{\frac{5\,{c^2}\,{{\left( 4\,{E^2} - m^2 \right) }^3}}
    {196608\,{E^7}\,{\sqrt{-0.04 + {E^2}}}\,m^2\,{{\pi }^2}}}
   + \frac{\left( 4\,{E^2} - m^2 \right)
       \left( {c^2}\,\left( 4\,{m^4} \right)  +
        {E^2}\,m^2\,\left( -16\,{c^2} + m^2\,\left( 0.96 + 3\,m^2 \right)
        \right)  \right)}{
        98304\,{E^5}\,{\sqrt{-0.04 + {E^2}}}\,
        {m^4}\,{{\pi }^2}}
 \nn\\
    &+&
  {\frac{{\sqrt{-0.04 + {E^2}}}\,\left( 4\,{E^2} - m^2 \right) \,
      \left( 256\,{E^8} - 256\,{E^6}\,m^2 - 16\,{E^2}\,{m^6} + {m^8} +
        {E^4}\,\left( 96\,{m^4} \right)  \right)
        }{393216\,{E^7}\,{m^4}\,{{\pi }^2}}}
  \nn\\
        &-&
  \frac{
  \left( 4\,{E^2} - m^2 \right) \,\left( 48\,{E^6}\,m^2 - 0.12\,{m^6}+
        {E^4}\,\left( 64\,{c^2} + \left( -1.92 - 24\,m^2 \right) \,m^2
        \right)\right)}{
        98304\,{E^5}\,{\sqrt{-0.04 + {E^2}}}\,
        {m^4}\,{{\pi }^2}}
        \nn\\
      &+&
  {\frac{{{\left( -4\,{E^2} + m^2 \right) }^2}\,
     \left(\, 48\,{E^6}\,m^2 - 0.12\,{m^6} +
        {E^4}\,\left( 64\,{c^2} + \left( -1.92 - 24\,m^2 \right)
        \,m^2 \right)\right)}{
        393216\,
      {E^7}\,{\sqrt{-0.04 + {E^2}}}\,{m^4}\,{{\pi }^2}}}
   \nn\\
        &+& {\frac{{{\left( -4\,{E^2} + m^2 \right) }^2}\,
        \left( 4\,{m^4}\,{c^2}  +
        {E^2}\,m^2\,\left( -16\,{c^2} + m^2\,\left( 0.96 + 3\,m^2 \right)
        \right)  \right) }{
        393216\,
      {E^7}\,{\sqrt{-0.04 + {E^2}}}\,{m^4}\,{{\pi }^2}}}
      \nn\\
      &-&
  {\frac{{{\left( -4\,{E^2} + m^2 \right) }^2}\,
      \left(  6\,{c^2}\,{m^4} + 0.08\,{m^6}
      + {\frac{c\,{m^6}}{2}} -
        32\,{E^6}\,\left( 2\,c + m^2 \right)
      + {E^4}\,m^2\,
        \left( 1.28 - 56\,c + 16\,m^2 \right)  \right)
        }{393216\,{E^7}\,{\sqrt{-0.04 + {E^2}}}\,{m^4}\,{{\pi }^2}}}
\nn\\
&-&  {\frac{{{\left( -4\,{E^2} + m^2 \right) }^2}\,
\left(\,
        {E^2}\,\left( 24\,{c^2}\,m^2 + \left( -0.64 - 2\,m^2 \right) \,{m^4}
        - 8\,{m^4}\,c  \right)  \right)
        }{393216\,{E^7}\,{\sqrt{-0.04 + {E^2}}}\,{m^4}\,{{\pi }^2}}},
\nn
\eea
where $c=\vec p_1 \cdot \vec k =|\vec p_1 ||\vec k|\cos \theta$.

\end{document}